\documentclass[10pt]{iopart}
\usepackage[OT1,T1]{fontenc}
\usepackage[latin1]{inputenc}
\usepackage{iopams} 
\usepackage{setstack}
\usepackage{bbm} 
\usepackage{undertilde} 
\newcommand{\di}{\mathrm{d}} 
\newcommand{\ou}[3]{{#1}{}^{#2}{}_{#3}} 
\newcommand{\uo}[3]{{#1}{}_{#2}{}^{#3}} 
\newcommand{\I}{\mathrm{i}} 
\newcommand{\E}{\mathrm{e}} 
\newcommand{\ellp}{{\ell_{\mathrm{P}}}} 
\newcommand{\CC}{\mathrm{cc.}} 
\usepackage{dsfont}
\usepackage[dvips]{graphicx}
\usepackage{color}
\usepackage{graphicx}
\newcommand{\qq}[1]{``#1''} 
\usepackage{hyperref}

\begin{document}
\title{Twistorial phase space for complex Ashtekar variables}
\author{Wolfgang M. Wieland}
\address{Centre de Physique Théorique,
   Campus de Luminy, Case 907,
   13288 Marseille, France, EU\footnote{Unité Mixte de Recherche (UMR 7332) du CNRS et de l'Université
d'Aix-Marseille et de l'Univ Sud Toulon Var. Unité affiliée à la
FRUMAM.}}
\eads{
\mailto{Wolfgang.Wieland@cpt.univ-mrs.fr}}
\date{July 2011}
\begin{abstract}
We generalise the $SU(2)$ spinor framework of twisted geometries developed by Dupuis, Freidel, Livine, Speziale and Tambornino to the Lorentzian case, that is the group $SL(2,\mathbb{C})$. We show that the 
phase space for complex valued Ashtekar variables on a spinnetwork graph can be decomposed in terms of twistorial variables. To every link there are two twistors---one to each boundary point---attached. The formalism provides a new derivation of the solution space of the simplicity constraints of loop quantum gravity. Key properties of the EPRL spinfoam model are perfectly recovered.
\end{abstract}
\section{Introduction}
In a series of pioneering articles \cite{twist, twist2, twist3, spinrep, twistcons} Dupuis, Freidel, Livine, Speziale and Tambornino developed a spinorial description of loop quantum gravity. They considered the case of $SU(2)$, which corresponds to the choice of real valued Ashtekar variables. In the present article we generalise this framework to the case of complex variables. The paper is tightly related to the triple  of articles that \cite{spezialetwist1, spezialetwist2, holsimpl} appeared just recently. 

The article is organised as follows. After shortly reviewing the phase space of general relativity in terms of connection variables, section \ref{II} discusses a certain truncation of that space. This truncation lives on an embedded graph, and provides the usual starting point for the program of loop quantisation. The reduced space is shown to be related to a number of copies of $T^\ast SL(2,\mathbb{C})$ equipped with a certain symplectic potential.

Section \ref{III} presents the first result of the article. We show that the truncated phase space allows for a spinorial decomposition. Every link of the graph is equipped with a pair of twistors, one attached to each of the two boundary points. The proof closely follows the lines of the $SU(2)$ case, developed by Freidel and Speziale in \cite{twist3}.

The second result is an application of the formalism. It is given in section \ref{IV}. In spinfoam gravity there are a number of simplicity constraints, matching the reality conditions of the Hamiltonian formulation. After having rewritten these constraint equations in terms of spinorial variables we canonically quantise. Nicely, the corresponding solution space coincides with the one found from the EPRL spinfoam model \cite{flppdspinfoam, LQGvertexfinite, lorentzvertam}.

Notation, conventions and derivations are collected in the appendix.
\section{The phase space for loop quantum gravity on a single link}\label{II}
\subsection{Smeared variables}
The Lagrangians for general relativity used within all modern approaches towards loop quantum gravity \cite{rovelli, thiemann} contain a parity breaking term proportional to the inverse Barbero--Immirzi parameter $\beta$. This parameter affects the symplectic potential of the theory: 
\begin{equation}
{\Theta}_{\mathrm{GR}}=\frac{\I\hbar}{2\ellp^2}\frac{\beta+\I}{\beta}\int_{t=\mathrm{const}.} P_{IJMN}\Sigma^{IJ}\wedge\mathbbm{d}A^{MN}+\CC\label{sympot}
\end{equation}
Here $P_{IJMN}$ is the selfdual projector, explicitly introduced in the appendix, $A^{IJ}$ denotes the $\mathfrak{so}(1,3)$ spin connection, and $\Sigma_{IJ}$ is the Plebanski 2-form:
\begin{equation}
\Sigma_{IJ}=e_I\wedge e_J,\label{Plbsk}
\end{equation}
where $e^I$ is the co-tetrad field. The integration goes over a $t=\mathrm{const}.$ hypersurface of initial data. Loop quantum gravity starts from a truncation of this phase space on a graph $\Gamma$ built from a finite number of piecewise analytic oriented paths or links $(\gamma_1,\dots,\gamma_L)$, possibly meeting at a certain number of nodes. Each of these links $\gamma$ is tranversally intersected by a dual surface \footnote{This surface is equipped with a natural orientation, that is the pair $U,V\in T_{\gamma(t_o)}f$ at the intersection point $\gamma(t_o)$ is said to be positively oriented provided the triple $(\dot\gamma(t_o),U,V)$ is positively oriented in $\partial M$.}, that is a face $f$. Both the connection and the Plebanski 2-Form can now naturally be smeared over these lower dimensional submanifolds, thereby obtaining the famous holonomy-flux variables:
\numparts
\begin{eqnarray}
g[f]&:=\mathrm{Pexp}\Big(\int_{\gamma} A\Big)\in SL(2,\mathbb{C}),\label{hlnmy}\\
\Pi[f]&:=\frac{\I\hbar}{2\ellp^2}\frac{\beta+\I}{\beta}\int_{q\in f}g_{(q\rightarrow p)}\Sigma_q g_{(q\rightarrow p)}^{-1}\in\mathfrak{sl}(2,\mathbb{C}).\label{flx}
\end{eqnarray}
\endnumparts
Here $\mathrm{Pexp}$ is the usual path-ordered exponential and $g_{(q\rightarrow p)}$ denotes a holonomy \footnote{The underlying system of paths decomposes into two parts, the first one lies inside $f$ mapping any $q\in f$ towards the intersection $f\cap\gamma$, whereas the next one goes from this intersection along $\gamma$ towards the initial point $\gamma(0)=p$.} parallely transporting any $q\in f$ towards the initial point $p=\gamma(0)$. In reference \cite{EEcomm} this construction is made extensively more explicit.
Moreover, in equation \eref{hlnmy} and \eref{flx} we implicitly used the canonical isomorphism \eref{lambdaiso} between $\mathfrak{sl}(2,\mathbb{C})$ and $\mathfrak{so}(1,3)$. 

We now take the Pauli spin matrices $\sigma_i=2\I\tau_i$ and define the selfdual components $\Pi[f]=\Pi_i[f]\tau^i$ of the momentum variable in order to compactly write the Poisson brackets for the left handed sector:
\numparts
\begin{eqnarray}
\big\{g[f],g[f^\prime]\big\}_{\mathrm{GR}} & = 0,\\
\big\{\Pi_i[f],g[f^\prime]\big\}_{\mathrm{GR}} & = \cases{+g[f]\tau_i,\;\mathrm{if:}\;\epsilon(f,f^\prime)=+1,\\-\tau_ig[f],\;\mathrm{if:}\;\epsilon(f,f^\prime)=-1,\\0,\;\mathrm{otherwise}.}\\
\big\{\Pi_i[f],\Pi_j[f^\prime]\big\}_{\mathrm{GR}} & = \delta_{ff^\prime}\uo{\epsilon}{ij}{m}\Pi_m[f].
\end{eqnarray}
\endnumparts
Here $\epsilon(f,f^\prime)$ denotes the relative orientation of the two surfaces, and $\delta_{ff^\prime}=|\epsilon(f,f^\prime)|$. The antiselfdual sector happens to be just the complex conjugate of the former, e.g.:
\begin{equation}
\big\{\bar\Pi_i[f],\ou{\bar{g}}{\bar{\mu}}{\bar\nu}[f]\big\}_{\mathrm{GR}}=\ou{\bar{g}}{\bar{\mu}}{\bar\alpha}[f]\ou{\bar\tau}{\bar\alpha}{\bar\nu i}.
\end{equation} 
Poisson brackets between variables of opposite chirality vanish. In the next section we will briefly sketch that the Poisson algebra of these smeared variables is naturally recovered from the cotangent bundle $T^\ast SL(2,\mathbb{C})$ equipped with a symplectic potential borrowed from the one of the continuum theory \eref{sympot}.
\subsection{The phase space T*SL(2,C)\label{BII}}
Consider the cotangent bundle $T^\ast SL(2,\mathbb{C})$ any point $(\sigma, g)$ of which consists of some group element $g\in SL(2,\mathbb{C})$ and a 1-Form $\sigma\in T^\ast_g SL(2,\mathbb{C})$ at $g$. If we define left invariant vector fields corresponding to both boost and rotations:
\numparts
\begin{eqnarray}
&X_i^\ell\big|_g=\frac{\di}{\di\varepsilon}\Big|_{\varepsilon=0}g\exp(\varepsilon\tau_i),\\
&Y_i^\ell\big|_g=\frac{\di}{\di\varepsilon}\Big|_{\varepsilon=0}g\exp(\varepsilon\I\tau_i),
\end{eqnarray}
\endnumparts
together with their complexified combinations:
\numparts
\begin{eqnarray}
&Z_i^\ell=\frac{1}{2}\Big(X_i^\ell-\I Y_i^\ell\Big),\\
&\bar{Z}_i^\ell=\frac{1}{2}\Big(X_i^\ell+\I Y_i^\ell\Big),
\end{eqnarray}
\endnumparts
we are able to perform a useful change of variables:
\begin{equation}
\eqalign{
T^\ast &SL(2,\mathbb{C})\rightarrow\mathfrak{sl}(2,\mathbb{C})\times SL(2,\mathbb{C}),\\
&(\sigma,g)\mapsto(\Sigma,g)
:\Sigma=(\ell_g^\ast\sigma)_{\mathds{1}}(Z_i)\tau^i,
}
\end{equation}
where $\ell_g^\ast$ denotes the differential map associated to left translation, i.e. $\ell_gg^\prime=gg^\prime$. Defining
\begin{equation}
\Theta_{\mathbf{P}}:=\frac{\beta+\I}{\I\beta}\mathrm{Tr}\big(\Sigma g^{-1}\di g\big)+\CC
\end{equation}
the cotangent bundle $T^\ast SL(2,\mathbb{C})$ inherits a natural symplectic structure from the continuous theory \eref{sympot}.
Notice the appearence of the Barbero--Immirzi parameter, implicitly showing that there is a 1-parameter family of mutually different symplectic potentials on $T^\ast SL(2,\mathbb{C})$ available. A fact which can be traced back to the existence of two independent Casimir operators for $SL(2,\mathbb{C})$.

Later it will prove necessary to work only with space- or timelike Lie algrebra elements, that is we exclude any $\Sigma:\mathrm{Tr}(\Sigma\Sigma)=0$, and define the truncated phase space:
\begin{equation}
\mathbf{P}=T^\ast SL(2,\mathbb{C})-\big\{\Sigma\in\mathfrak{sl}(2,\mathbb{C})\big|\mathrm{Tr}(\Sigma\Sigma)=0\big\}.
\end{equation}
From the symplectic potential we deduce the corresponding Poisson algebra. First of all we find that the Poisson brackets between matrix entries of group elements vanish trivially:
\begin{equation}
\big\{\ou{g}{\alpha}{\beta},\ou{g}{\mu}{\nu}\big\}_{\mathbf{P}}=0\label{ggbrace}
\end{equation}
A straightforward calculation reveals those Poisson brackets that contain momentum variables:
\numparts
\begin{eqnarray}
\big\{&\Pi_i,\ou{g}{\alpha}{\beta}\big\}_{{\mathbf{P}}}  =  \ou{(g\tau_i)}{\alpha}{\beta},\label{ppbka}\\
\big\{&\Pi_i,\Pi_j\big\}_{\mathbf{P}}  =  \uo{\epsilon}{ij}{m}\Pi_m,\label{ppbkb}
\end{eqnarray}
\endnumparts
where we have introduced the abbreviation:
\begin{equation}
\Pi=-\frac{1}{2}\frac{\beta+\I}{\I\beta}\Sigma.\label{PSrel}
\end{equation}
Notice also that for the complex conjugate variables, that is the sector of opposite chirality the Poisson brackets remain qualitatively unchanged:
\numparts
\begin{eqnarray}
\big\{&\bar\Pi_i,\ou{\bar{g}}{\bar\alpha}{\bar\beta}\big\}_{\mathbf{P}}  =  \ou{(\bar{g}\bar{\tau}_i)}{\bar\alpha}{\bar\beta},\label{ppbbka}\\
\big\{&\bar\Pi_i,\bar\Pi_j\big\}_{\mathbf{P}}  =  \uo{\epsilon}{ij}{m}\bar\Pi_m.\label{ppbbkb}
\end{eqnarray}
\endnumparts
All Poisson brackets between variables of mutually different chirality vanish.
Finally, we may also wish to introduce the left invariant Lie algebra element:
\begin{equation}
\utilde{\Pi}=-g\Pi g^{-1}\label{armatch},
\end{equation}
for which the Poisson brackets turn out to be:
\numparts
\begin{eqnarray}
\big\{\Pi_i,\utilde{\Pi}_j\big\}_{\mathbf{P}} & = 0,\label{pptbka}\\
\big\{\utilde{\Pi}_i,\ou{g}{\alpha}{\beta}\big\}_{\mathbf{P}} & = -\ou{(\tau_ig)}{\alpha}{\beta},\label{pptbkb}\\
\big\{\utilde{\Pi}_i,\utilde{\Pi}_j\big\}_{\mathbf{P}} & =\uo{\epsilon}{ij}{m}\utilde{\Pi}_m.\label{pptbkc}
\end{eqnarray}
\endnumparts
For a single link we interpret any point $(\Sigma, g)\in\mathbf{P}$ as follows. The rescaled momentum $\Pi$ defined as in \eref{PSrel} corresponds to the flux $\Pi[f]$ through the surface dual to the link, whereas $\utilde{\Pi}$ is related to $\Pi[f^{-1}]$, i.e. the momentum smeared over the oppositely oriented surface. The group element $g$ in $(\Sigma, g)\in\mathbf{P}$ is attached to the whole link and represents the holonomy $g[f]$ from the source $p$ towards the target point $\utilde{p}$. This correspondence can trivially be generalised to a graph containing $L$ links, where $\mathbf{P}$ is simply replaced by a number of $L$ copies of $\mathbf{P}$. 
\section{Twistorial decomposition}\label{III}
Having introduced the phase space ${\mathbf{P}}$ of interest, we are now ready to go into the main part of this article, where a spinorial (or rather twistorial) decomposition will be developed.
\subsection{Twistorial phase space on a link}
A twistor $Z\in\mathbb{T}$ \cite{pentwist2, pentwist1} is a bispinor \cite{penroserindler1}
\begin{equation}
Z=(\omega^\mu,\bar{\pi}_{\bar{\nu}})\in\mathbb{C}^2\oplus\big(\bar{\mathbb{C}}^2\big)^\ast=\mathbb{T},
\end{equation}
the two components $\omega$, $\bar{\pi}$ of which are elements of $\mathbb{C}^2$ and its complex conjugate dual vector space, i.e. $\big(\bar{\mathbb{C}}^2\big)^\ast$. The two respective parts transform according to
\numparts
\begin{eqnarray}
\omega^\mu&\stackrel{g}{\longrightarrow}+\ou{g}{\mu}{\nu}\omega^\nu,\\
\bar{\pi}_{\bar{\mu}}&\stackrel{g}{\longrightarrow}-\uo{\bar{g}}{\bar{\mu}}{\bar{\nu}}\bar{\pi}_{\bar{\nu}}\label{dualtrafo},
\end{eqnarray}
\endnumparts
under the action of the $SL(2,\mathbb{C})$ group. 
On the space ${\mathbb{T}}$ of twistors there is an invariant symplectic structure available, which is entirely defined by the only nonvanishing Poisson brackets
\numparts
\begin{eqnarray}
\big\{\pi_\nu,\omega^\mu\big\}_{\mathbb{T}} & = \delta^\mu_\nu,\\
\big\{\bar{\pi}_{\bar{\nu}},\bar{\omega}^{\bar{\mu}}\big\}_{\mathbb{T}} & = \bar{\delta}^{\bar{\mu}}_{\bar{\nu}},
\end{eqnarray}\label{spinpoiss}
\endnumparts
naturally generated by the symplectic potential $\Theta_{\mathbb{T}}=\pi_\mu\di\omega^\mu+\bar{\pi}_{\bar\mu}\di\bar{\omega}^{\bar\mu}$, and the corresponding real valued 2-form $\Omega_{\mathbb{T}}=\di\Theta_{\mathbb{T}}$. Note that this symplectic structure differs from the definitions commonly used in the literature by a trivial rescaling transformation $\bar{\pi}\rightarrow\I\bar{\pi}$.

The tensor product $\pi^\alpha\omega^\beta$ of the two left handed phase space variables transforms under a reducible transformation of $SL(2,\mathbb{C})$, the two irreducible parts are given by:
\numparts
\begin{eqnarray}
H&=\pi_\mu\omega^\mu,\label{hel}\\
\Pi_i&=\tau_{\alpha\beta i}\pi^\alpha\omega^\beta.\label{ang}
\end{eqnarray}
\endnumparts
These are certainly not the only interesting spinorial bilinears we can think of, in fact one can construct e.g.
\begin{equation}
\Xi_I=\I\ou{\sigma}{\alpha\bar\beta}{I}\pi_\alpha\bar{\omega}_{\bar\beta}+\CC
\end{equation}
But for the purpose of this article we need just \eref{hel} and \eref{ang}. It is quite obvious to show that these variables obey the following Poisson commutation relations:
\numparts
\begin{eqnarray}
\big\{\Pi_i,\Pi_j\big\}_{\mathbb{T}} & = \uo{\epsilon}{ij}{m}\Pi_m,\label{ppspin}\\
\big\{\Pi_{i},H\big\}_{\mathbb{T}} & = 0.
\end{eqnarray}
\endnumparts
Again all Poisson brackets between the two sectors of opposite chirality, e.g. $\{\Pi_i,\bar{H}\}_{\mathbb{T}}=0$ vanish trivially. Equation \eref{ppspin} tells us that the Hamiltonian vector field associated to $\Pi_i$ generates Lorentz transformations on phase space, whereas $H$ is responsible for infinitesimal scaling transformations, to be a little more precise we find that:
\numparts
\begin{eqnarray}
\big\{\Pi_i,\pi^\alpha\big\}_{\mathbb{T}}&=-\ou{\tau}{\alpha}{\beta i}\pi^\beta,\label{genactna}\\
\big\{\Pi_i,\omega^\alpha\big\}_{\mathbb{T}}&=-\ou{\tau}{\alpha}{\beta i}\omega^\beta,\label{genactnb}
\end{eqnarray}
\endnumparts
together with:
\numparts
\begin{eqnarray}
\big\{H,\pi^\alpha\big\}_{\mathbb{T}}&=-\pi^\alpha,\\
\big\{H,\omega^\alpha\big\}_{\mathbb{T}}&=+\omega^\alpha.
\end{eqnarray}\label{scaltraf}
\endnumparts
In order to establish a spinorial decomposition of $T^\ast SL(2,\mathbb{C})$, that is the phase space attached to each of the links of the graph, let us consider a pair of twistors,
\begin{equation}
(\utilde{Z},Z)=(\utilde{\omega}^\mu,\utilde{\bar\pi}_{\bar\mu},\omega^\mu,\bar{\pi}_{\bar\mu}).
\end{equation}
equipped with the natural Poisson bracket already introduced:
\begin{equation}
\{\pi_\mu,\omega^\nu\}_{\mathbb{P}}=\{\utilde{\pi}_\mu,\utilde{\omega}^\nu\}_{\mathbb{P}}=\delta^\nu_\mu,
\end{equation}
Furthermore for our construction to work null elements
\begin{equation}
\mathbb{T}_0:=\big\{(\utilde{Z},Z)\big|\pi_\mu\omega^\mu=0,\;\mathrm{or}\;\,\utilde{\pi}_\mu\utilde{\omega}^\mu=0\big\}
\end{equation}
need to be removed, and we use the symbol
\begin{equation}
\mathbb{P}:=\mathbb{T}\times\mathbb{T}-\mathbb{T}_0
\end{equation}
in order to refer to the phase space so defined. The relation to the geometry of the graph is indicated by our notation. Variables marked with a tilde (e.g. $\utilde{Z}$) refer to the final point, whereas the twistor $Z$ is attached to the initial point.
\subsection{Twistorial decomposition of the phase space variables}
Let us now show how the phase space parametrised by $\Sigma\in\mathfrak{sl}(2,\mathbb{C})$ and $g\in SL(2,\mathbb{C})$ decomposes in terms of our pair of twistors.
Consider first the following matrix:
\begin{equation}
\ou{g(\utilde{Z},Z)}{\alpha}{\beta}\equiv\ou{g}{\alpha}{\beta}=\frac{\utilde{\pi}^\alpha\pi_\beta+\utilde{\omega}^\alpha\omega_\beta}{\sqrt{\utilde{\pi}_\mu\utilde{\omega}^\mu\pi_\nu\omega^\nu}}.\label{gdef}
\end{equation}
Quite obviously $g\in SL(2,\mathbb{C})$, furthermore by setting $\pi_\alpha=\delta_\alpha^o\equiv o_\alpha$, and $\omega_\alpha=\delta_\alpha^1\equiv\iota^\alpha$, together with $\utilde{\pi}^\beta={{a}\choose{b}}^\beta$ and $\utilde{\omega}^\beta={c\choose{d}}^\beta$ we get
\begin{equation}
g(\utilde{Z},Z)=\frac{\pmatrix{a & c \cr b & d}}{\sqrt{ad-bc}},
\end{equation}
and can hence immediately deduce that any $SL(2,\mathbb{C})$ element can be written in the form of \eref{gdef}. Interested to decompose all of phase space $T^\ast SL(2,\mathbb{C})\simeq\mathfrak{sl}(2,\mathbb{C})\times SL(2,\mathbb{C})$ into spinorial variables, we have just achieved half of this task. Consider the following ansatz for the Lie algebra part:
\numparts
\begin{eqnarray}
\Pi^{\alpha\beta}&=\frac{1}{4}\big(\pi^\alpha\omega^\beta+\pi^\beta\omega^\alpha\big),\label{liedef}\\
\utilde{\Pi}^{\alpha\beta}&=\frac{1}{4}\big(\utilde{\pi}^\alpha\utilde{\omega}^\beta+\utilde{\pi}^\beta\utilde{\omega}^\alpha\big).\label{liedef2}
\end{eqnarray}\label{liedef3}
\endnumparts
And \eref{PSrel} provides the relation between $\Pi$ and $\Sigma$. Since $\ou{\Pi}{\alpha}{\alpha}=0$ we truly found a Lie algebra element, in addition we may easily convince ourselves that the decomposition captures all Lie algebra elements, except of those being null:
\begin{equation}
\Pi^{\alpha\beta}\Pi_{\alpha\beta}=0.
\end{equation}
This restriction comes from the fact that the hypersurface $\pi_\mu\omega^\mu=0$ had to be removed in order to keep \eref{gdef} well defined. Concerning the parametrisation defined by \eref{liedef} and \eref{gdef} it is still an open question whether it reaches any \emph{pair} $(\Sigma,g)\in\mathbf{P}$. But this is indeed the case.

In order to prove this, let us consider first a \qq{twisted rotation} introduced by Livine and Tambornino for the case of $SU(2)$ in \cite{spinrep}. Be $G$ some $GL(2,\mathbb{C})$ group element, such that we can define the following transformations:
\numparts
\begin{eqnarray}
\omega^\alpha  \longrightarrow \ou{G}{\alpha}{\beta}\omega^\beta,&\quad \utilde{\omega}^\alpha  \longrightarrow \ou{(gGg^{-1})}{\alpha}{\beta}\utilde{\omega}^\beta,\label{twistrota}\\
\pi^\alpha  \longrightarrow \ou{G}{\alpha}{\beta}\pi^\beta,&\quad \utilde{\pi}^\alpha  \longrightarrow \ou{(gGg^{-1})}{\alpha}{\beta}\utilde{\pi}^\beta.\label{twistrotb}
\end{eqnarray}
\endnumparts
Where $g$ equals the group element \eref{gdef} constructed from the twistorial variables. In the appendix we will prove \eref{twistrotprf} that this transformation actually leaves the \qq{holonomy} as defined by \eref{gdef} invariant. However the Lie algebra element gets transformed non trivially:
\begin{equation}
\Pi^{\alpha\beta}\longrightarrow \ou{G}{\alpha}{\mu}\ou{G}{\beta}{\nu}\Pi^{\mu\nu}.
\end{equation}
Consider now our example for which $\pi_\alpha=o_\alpha$, and $\omega_\alpha=\iota_\alpha$. Be $p_\mu$ and $z^\mu$ another pair of spinors, which parametrise the desired Lie algebra element $\ou{\Pi}{\alpha}{\beta}$ according to $4\Pi^{\alpha\beta}=z^\alpha p^\beta+z^\beta p^\alpha$, and $p_\mu z^\mu\neq 0$ be fulfilled. The linear map $G:\mathbb{C}^2\rightarrow\mathbb{C}^2$ entirely defined by its action onto the basis elements $(o,\iota)$ according to
\begin{equation}
G:(o^\alpha,\iota^\alpha)\mapsto(p^\alpha,z^\alpha),
\end{equation}
is a proper element of $GL(2,\mathbb{C})$. In other words; by the use of a twisted rotation we can---for any pair $(\Pi,g)$ consisting of a non singular $\mathfrak{sl}(2,\mathbb{C})$ element $\Pi^{\alpha\beta}\Pi_{\alpha\beta}\neq 0$ and a group element $g\in SL(2,\mathbb{C})$---always find a pair $(Z,\utilde{Z})$ of twistors such that \eref{gdef} and \eref{liedef} are fulfilled.

But for any given point $(\Sigma,g)\in\mathbf{P}$ the pair $(\utilde{Z},Z)$ of twistors is certainly not uniquely defined. It is not very hard to prove, in fact, that all pairs $(\utilde{Z},Z)$ that correspond to the very same point $(\Sigma,g)$ in $\mathbf{P}$ form a complex \qq{ray} $(\utilde{Z}(z),Z(z))$, parametrised by $z\in\mathbb{C}$:
\numparts
\begin{eqnarray}
\pi^\alpha(z)=\E^{-z}\pi^\alpha, &\quad \omega^\alpha(z)=\E^{z}\omega^\alpha,\label{confa}\\
\utilde{\pi}^\alpha(z)=\E^{z}\utilde{\pi}^\alpha, &\quad \utilde{\omega}^\alpha(z)=\E^{-z}\utilde{\omega}^\alpha.\label{confb}
\end{eqnarray}
\endnumparts
Next, we may observe that the pair $(\Pi,g)$ is unchanged if $\utilde{Z}$ is replaced by $\E^z\utilde{Z}$. But there is equation \eref{armatch} telling us that $\Pi$ and $\utilde{\Pi}$ cannot be chosen independently. This constraint leads in fact to the \qq{area matching condition} \cite{twist}
\begin{equation}
C=H-\utilde{H}=\pi_\mu\omega^\mu-\utilde{\pi}_\mu\utilde{\omega}^\mu=0.\label{armatch2}
\end{equation}
The proof, purely algebraically and an exercise in index manipulations may be found in appendix B. The residual symmetries of the pair of twistors that leave  any $(\Sigma,g)\in\mathbf{P}$ invariant are then the scaling transformations given in (\ref{confa}\emph{-b}), together with the map
\begin{equation}
(\pi,\omega,\utilde{\pi},\utilde{\omega})\longmapsto(\omega,\pi,\utilde{\omega},\utilde{\pi})\label{distrans}
\end{equation}
exchanging the two respective parts of the bispinors.
\subsection{Poisson brackets and symplectic reduction}
Up to here we have just seen that any point in $\mathbf{P}$, that is the phase space attached to a link, can be parametrised by a pair of twistors provided the constraint equation \eref{armatch2} holds. What we now show is that the Poisson commutation relations are equally well satisfied.

Defining the selfdual components of the momentum variables as usual, e.g. $\Pi_i=2{\tau}_{\alpha\beta i}{\Pi}^{\alpha\beta}$, we arrive at an already very familiar symplectic structure. In fact \eref{ppspin} together with (\ref{genactna}\emph{-b})  and \eref{gdef} imply that all Poisson bracket containing the canonical moments $\Pi_i$, $\utilde{\Pi}_i$, $\bar\Pi_i$ and $\utilde{\bar\Pi}_i$ trivially reproduce the corresponding Poisson brackets (\ref{ppbka}\emph{-b}, \ref{ppbbka}\emph{-b}, \ref{pptbka}\emph{-c}) on $\mathbf{P}$. E.g.:
\begin{equation}
\big\{\Pi_i,\ou{g}{\alpha}{\beta}\big\}_{\mathbb{P}}=\ou{(g\tau_i)}{\alpha}{\beta}.
\end{equation}
These equations are fulfilled on all of phase space, for the Poisson brackets between group elements \eref{gdef} the situation is different. In fact 
\begin{equation}
\big\{\ou{g}{\alpha}{\beta},\ou{g}{\mu}{\nu}\big\}_{\mathbb{P}}\stackrel{\mathrm{in\;general}}{\neq}0
\end{equation}
does not generally vanish, and we should worry if we could implement \eref{ggbrace} in terms of spinorial variables. However there is the additional constraint \eref{armatch2} to be fulfilled. In the appendix we will show that on the constraint hypersurface the vanishing of \eref{ggbrace} is fully recovered:
\begin{equation}
\big\{\ou{g}{\alpha}{\beta},\ou{g}{\mu}{\nu}\big\}_{\mathbb{P}}\big|_{C=0}=0.
\end{equation}
We are now ready to clarify the relation between $\mathbf{P}$ and the twistorial phase space $\mathbb{P}$. The Hamiltonian vector field $\mathfrak{X}_C=\{C,\cdot\}_{\mathbb{P}}$ generates a flow on phase space tangential to the constraint hypersurface $C=0$. From \eref{scaltraf} we deduce for any $z\in\mathbb{C}$ its action to be:
\numparts
\begin{eqnarray}
\exp\big(z\mathfrak{X}_C\big)\pi^\alpha&=\E^{-z}\pi^\alpha,&\quad \exp\big(z\mathfrak{X}_C\big)\omega^\alpha=\E^{z}\omega^\alpha,\\
\exp\big(z\mathfrak{X}_C\big)\utilde{\pi}^\alpha&=\E^{z}\utilde{\pi}^\alpha,&\quad \exp\big(z\mathfrak{X}_C\big)\utilde{\omega}^\alpha=\E^{-z}\utilde{\omega}^\alpha,
\end{eqnarray}
\endnumparts
which obviously coincides with (\ref{confa}\emph{-b}). This transformation leaves the pair $(\Pi, g)$ unchanged, that is $\{C,g\}_{\mathbb{P}}=\{C,\Pi\}_{\mathbb{P}}=0$. The symplectic potential $\Theta_{\mathbb{T}}=\pi_\mu\di\omega^\mu+\utilde{\pi}_\mu\di\utilde{\omega}^\mu+\CC$ is equally invariant under this flow. We can thus perform a symplectic reduction 
\begin{equation}
\boxed{\mathbb{P}/\!\!/_{C}=\mathbf{P}}
\end{equation}
obtaining\footnote{In fact one should also \qq{divide} here by the discrete transformation \eref{distrans}.} the original phase space introduced in section \ref{BII}.  By this symplectic reduction, points on the constraint hypersurface lying on the same orbit generated by the action of $\mathfrak{X}_C$ are identified. Notice also that the phase space dimensions are correctly reduced, $\mathbb{P}$ has $4+4$ complex degrees of freedom, the constraint $C=0$ removes one of them. The identification of the gauge orbits generated by the action of $C$ removes another complex dimension. We are thus left with 6 complex degrees of freedom perfectly matching the 12 real dimensions of $T^\ast SL(2,\mathbb{C})\simeq\mathfrak{sl}(2,\mathbb{C})\times SL(2,\mathbb{C})$.
\section{Simplicity constraints}\label{IV}
\subsection{Classical treatment}
In this section we wish to illustrate the computational power of the twistorial formalism just developed. We do this by deriving two equations key to the definition of the EPRL spinfoam model \cite{flppdspinfoam, LQGvertexfinite, lorentzvertam}.

General relativity can be described in terms of a topological theory the field content of which 
is constrained by a number of simplicity constraints \cite{physbound, covol, genspinfoam}. Be there a quantisation of both the topological theory and the additional simplicity constraints, a clean definition of the gravitational path integral seems feasible. This is of course the key idea the EPRL spinfoam model is built on. 

Where do these simplicity constraints actually come from? 
In spinfoam gravity we treat the Plebanski 2-form \eref{Plbsk} as a fundamental quantity. But in general relativity this is a derived object: Given some typical $\mathfrak{so}(1,3)$ valued 2-form $\Sigma_{IJ}$, there is not necessarily a co-tetrad associated to it. Additional constraint equations are needed, in fact the \emph{linear simplicity constraints} (together with \emph{Gauß's law}) restrict the phase space to those $\Sigma_{IJ}$ for which a co-tetrad can always be found. In the canonical framework the simplicity constraints naturally appear as \emph{reality conditions} on the momentum variable \cite{komplex1}. For the truncated theory living on a fixed graph these constraints require at any node $p$ the existence of an internal normal $n^I$ fulfilling
\begin{equation}
\Sigma_{IJ}[f]n^J=0,\label{simpl}
\end{equation} 
for all faces $f$ adjacent to $p$. 
By the use of spinorial variables equation  \eref{simpl} turns into
\begin{equation}
\Sigma_{\alpha\beta}[f]\bar\epsilon_{\bar\alpha\bar\beta}n^{\beta\bar\beta}+\CC=0,\label{simplx}
\end{equation}
where $\Sigma_{\alpha\beta}[f]$ and $\bar\Sigma_{\bar\alpha\bar\beta}[f]$ refer to the self- and antiselfdual parts of $\Sigma_{IJ}[f]$, and $n^{\alpha\bar\alpha}=\ou{\sigma}{\alpha\bar\alpha}{I}n^I$ equals the internal normal contracted with the Pauli matrices \eref{Paulm}.
Inserting the momentum conjugate of the connection \eref{simplx} takes the following form
\begin{equation}
\frac{\I\beta}{\beta+\I}\Pi_{\alpha\beta}[f]\bar\epsilon_{\bar\alpha\bar\beta}n^{\beta\bar\beta}+\CC=0\label{spinsimpl}.
\end{equation}
In terms of our twistorial variables this leads us to
\begin{equation}
\frac{\I\beta}{\beta+\I}\big(\omega_\alpha\pi_\beta+\omega_\beta\pi_\alpha\big)\bar\epsilon_{\bar\alpha\bar\beta}n^{\beta\bar\beta}+\CC=0.\label{lincon1}
\end{equation}
We are interested in the case most seriously studied 
in the literature, that is we choose the internal normal to be timelike; $n^{\alpha\bar\alpha}n_{\alpha\bar\alpha}=+2$. If we now contract \eref{spinsimpl} by the matrix $\omega^\alpha\bar{\omega}^{\bar\alpha}$ we find the following equation:
\begin{equation}
F_1=\frac{\I}{\beta+\I}\omega^\alpha\pi_\alpha+\CC=0.\label{simpl1}
\end{equation}
Contracting \eref{spinsimpl} by $n^{\alpha\bar\mu}\bar\omega_{\bar\mu}\bar{\omega}^{\bar\alpha}$ we get another constraint:
\begin{equation}
F_2=n^{\alpha\bar\beta}\pi_\alpha\bar{\omega}_{\bar\beta}=0.\label{simpl2}
\end{equation} 
From the contraction by $n^{\alpha\bar\mu}n^{\nu\bar\alpha}\bar\omega_{\bar\mu}\omega_{\nu}$ we would again get $F_1=0$. But the pair $(\omega^\alpha,n^{\alpha\bar\alpha}\bar\omega_{\bar\alpha})$ is a complete basis in $\mathbb{C}^2$ and therefore both \eref{simpl1} and \eref{simpl2} are actually already sufficient in order to prove that \eref{spinsimpl} is satisfied.

If we now study the respective Poisson brackets, we first get
\numparts
\begin{eqnarray}
\big\{F_1,F_2\big\}_{\mathbb{T}}&=-\frac{2\I\beta}{\beta^2+1}F_2,\\
\big\{F_1,\bar{F}_2\big\}_{\mathbb{T}}&=+\frac{2\I\beta}{\beta^2+1}\bar{F}_2.
\end{eqnarray}\label{F12}
\endnumparts
Therefore the constraint $F_1$ is of first class, and we expect that the corresponding quantum operator can be imposed strongly, that is it should annihilate physical states. For $F_2$ the situation is different, in fact we get
\begin{equation}
\big\{F_2,\bar{F}_2\big\}_{\mathbb{T}}=\pi_\alpha\omega^\alpha-\bar{\pi}_{\bar\alpha}\bar\omega^{\bar\alpha}.
\end{equation}
It is of second class, and there are several different ways to deal with this situation. Here Thiemann's master constraint approach, originally developed in \cite{Thiemann2003-MC} for full loop quantum gravity, will prove considerably simple and useful.
Let us define this constraint just as the square modulus
\begin{equation}
\boldsymbol{\mathsf{M}}=\bar{F_2}F_2,
\end{equation}
the vanishing of which is trivially equivalent to the vanishing of $F_2$. But in contrast to the case of $F_1$ and $F_2$ the corresponding constraint algebra is of first class:
\begin{equation}
\big\{\boldsymbol{\mathsf{M}},F_1\big\}_{\mathbb{T}}=0.
\end{equation}
In quantum theory the equations $\boldsymbol{\mathsf{M}}=0$ and $F_1=0$ can be imposed strongly, and should therefore be favoured over the second class constraints $F_1$ and $F_2$.

The following equivalent form, explicitly derived in an additional appendix, will prove increasingly more useful:
\begin{equation}
\boldsymbol{\mathsf{M}}=-\frac{1}{2}\bar{\omega}^{\bar\mu}\bar{\pi}_{\bar\mu}\pi_\mu\omega^\mu-\frac{1}{2}\big(L^2-K^2\big)+L^2.
\end{equation}
Here we have implicitly chosen a Lorentz frame aligning $n^I$ to $\delta^I_0$. Moreover $K_i$ and $L_i$ are the boost and rotation components of the momentum variable $2\Pi_i=-L_i-\I K_i$, and $L^2=L_iL^i$. 
\subsection{Gauß's law}
The simplicity constraints guarantee the existence of a tetrad only provided the geometry is non-degenerate ($\epsilon^{IJLM}\Sigma_{IJ}\wedge\Sigma_{LM}\neq0$) and the torsion-free condition (Gauß's law) $D\wedge \Sigma_{IJ}=0$ holds. By Stoke's theorem its smeared version over a three dimensional region---say a tetrahedron---equals the flux through its boundary. But we already gave the twistorial decomposition of the self- and antiselfdual parts of the smeared fluxes in \eref{liedef}. Be $f^1,\dots, f^4$ the four faces of the tetrahedron.  Assign a pair $(\omega^\mu(i),\bar{\pi}_{\bar\mu}(i))$ to each of the four faces $f^i$ dual to the corresponding \qq{half-links}. Provided all faces are oriented outwardly pointing, we find the twistorial decomposition of the selfdual $\mathfrak{sl}(2,\mathbb{C})$ fluxes to be
\begin{equation}
\Pi^{\alpha\beta}[f^i]=\frac{1}{4}\big(\pi^\alpha(i)\omega^\beta(i)+\pi^\beta(i)\omega^\alpha(i)\big)\label{fluxi}.
\end{equation}
And again the antiselfdual fluxes are just the complex conjugate of the former. Gauß's law then turns into
\begin{equation}
\sum_{i=1}^4\Pi^{\alpha\beta}[f^i]=0=\sum_{i=1}^4\bar\Pi^{\bar\alpha\bar\beta}[f^i].
\end{equation}
The Poisson commutation relations between the various components of the Gauß law can immediately be referred from (\ref{pptbka}\emph{-c}). 

In this article we wish to examine the twistorial structure on basically one single link. But imposing the Gauß constraint requires to know the adjacency relations of all different links of the spinnetwork graph. A more detailed discussion of the Gauß constraint would thus overstep the scope of this paper. The linear simplicity constraint on the other hand, being naturally smeared over one single face, can perfectly be solved on each link separately, and thus nicely fits into this article.
\subsection{Quantum theory}
From the canonical Poisson commutation relations \eref{spinpoiss} we can easily deduce the quantisation of the momentum variable
\numparts
\begin{eqnarray}
\bar\pi_{\bar\alpha}&\stackrel{\mathrm{quantisation}}{\longrightarrow} \frac{1}{\I}\frac{\partial}{\partial \bar\omega^{\bar\alpha}},\\
\pi_{\alpha}&\stackrel{\mathrm{quantisation}}{\longrightarrow} \frac{1}{\I}\frac{\partial}{\partial \omega^\alpha}.
\end{eqnarray}
\endnumparts
Where we have secretly assumed that a quantum state is given as a complex valued function of the \qq{configuration} variable $\omega$. We may also wish to think of a state $f$ as a square integrable function $f\in L^2(\mathbb{C}^2, \di\omega_\mu\wedge\di\omega^\mu\wedge\CC)$, a restriction which will however soon turn out to be rather inconvenient.

We will now find the quantisation of $\boldsymbol{\mathsf{M}}=F_1=0$, together with the states that are annihilated by the constraints. 
Choosing a \qq{normal ordering}, that is
\begin{equation}
\pi_\mu\omega^\mu\stackrel{\mathrm{quantisation}}{\longrightarrow} \frac{1}{2\I}\big(\omega^\mu\frac{\partial}{\partial\omega^\mu}+\frac{\partial}{\partial\omega^\mu}\omega^\mu\big),
\end{equation} 
the quantisation of the $F_1$ constraint becomes obvious:
\begin{equation}
\widehat{F}_1:=\frac{1}{\beta^2+1}\Big[(\beta-\I)\omega^\alpha\frac{\partial}{\partial\omega^\alpha}-(\beta+\I)\bar\omega^{\bar\alpha}\frac{\partial}{\partial\bar\omega^{\bar\alpha}}-2\I\Big].
\end{equation}
Notice the appearance of Euler homogeneity operators, which naturally suggests to find solutions of the corresponding constraint equation in terms of homogenous functions of two complex variables. Following the terminology of e.g. \cite{unrepsl, lorentzvertam} we call a function $f:\mathbb{C}^2\rightarrow\mathbb{C}^2$ homogenous of degree $(a,b)\in\mathbb{C}^2$ provided that
\begin{equation}
\forall \lambda\in\mathbb{C}:f(\lambda\omega^\alpha)=\lambda^a\bar\lambda^bf(\omega^\alpha),
\end{equation}
From the two 1-parameter families $\lambda_t=\E^t$ and $\lambda_t=\E^{\I t}$ we find the action of the homogeneity operators on functions of fixed degree $(a,b)$ to be:
\numparts
\begin{eqnarray}
\omega^\mu\frac{\partial}{\partial \omega^\mu}f=af,\\
\bar\omega^{\bar\mu}\frac{\partial}{\partial \bar\omega^{\bar\mu}}f=bf.
\end{eqnarray}
\endnumparts
For any $f$ of degree $(a,b)$ the constraint turns into the eigenvalue equation
\begin{equation}
\widehat{F}_1f=\frac{\beta(a-b)-\I(a+b)-2\I}{\beta^2+1}f\stackrel{!}{=}0.\label{eigen}
\end{equation}
Let us further restrict ourselves to irreducible unitary representations \cite{gelminshap, unrepsl} of the Lorentz group, for which the degrees of homogeneity are parametrised by $a=-j_o-1+\I\rho$ and $b=j_o-1+\I\rho$, where $2j_o\in\mathbb{N}_0$ and $\rho\in\mathbb{R}$. This choice seems reasonable, since it will naturally provide us with an $SL(2,\mathbb{C})$ invariant inner product on the solution space of the constraint equation. Inserting this parametrisation into the eigenvalue equation \eref{eigen} we find that the continous label is related to the discrete via the Barbero--Immirzi parameter:
\begin{equation}
\boxed{-\beta j_o+\rho=0
}
\end{equation}
This is one of the two constraint equations needed to define the EPRL vertex amplitude. The other one is recovered as follows. The quantisation of $\widehat{F}_2$ is unambiguous, indeed we have
\begin{equation}
\widehat{F}_2:=-\I n^{\alpha\bar\alpha}\bar\omega_{\bar\alpha}\frac{\partial}{\partial \omega^\alpha},\label{F2}
\end{equation}
but for the quantisation of the master constraint we have to choose an ordering:
\begin{equation}
\boldsymbol{\mathsf{\widehat{M}}}:=\widehat{F}_2^\dagger\widehat{F}_2=\frac{1}{2}\omega^\mu\frac{\partial}{\partial\omega^\mu}\frac{\partial}{\partial \bar\omega^{\bar\mu}}\bar\omega^{\bar\mu}-\frac{1}{2}\big(\widehat{L}^2-\widehat{K}^2\big)+\widehat{L}^2,
\end{equation}
where $\widehat{L}^2=\widehat{L}_i\widehat{L}^i$ is the squared angular momentum operator, $\widehat{K}_i$ generates boosts along the $i$-th direction, the Hermitian conjugate is taken with respect to the usual $L^2$ inner product on $\mathbb{C}^2$, and the proof follows the lines of \eref{Mcons}.

Consider now the canonical basis vectors $f^{(\rho,j_o)}_{j,m}$, which diagonalise \cite{ruhl} the operators $\widehat{L}^2-\widehat{K}^2$, $\widehat{L}_i\widehat{K}^i$, $\widehat{L}^2$ and $\widehat{L}_3$. The corresponding eigenvalues can be found in the appendix, and the spin quantum numbers $j=j_o,j_o+1,\dots$ and $m=-j,\dots, j$ refer to the $SU(2)$ invariant subspaces of the $SL(2,\mathbb{C})$ irreducible unitary representation space. In this basis the master constraint becomes diagonal:
\begin{eqnarray}
\nonumber\boldsymbol{\mathsf{\widehat{M}}}f^{(\rho,j_o)}_{j,m}&=\frac{1}{2}\Big[a(b+2)-(j_o^2-1-\rho^2)+\\
\nonumber&\qquad+2j(j+1)\Big]f^{(\rho,j_o)}_{j,m}=\\
\nonumber&=\frac{1}{2}\Big[-(j_o+1 )^2-(j_o+1)(j_o-1)+\\
&\qquad+2j(j+1)\Big]f^{(\rho,j_o)}_{j,m}\stackrel{!}{=}0.
\end{eqnarray}
There is just one solution possible:
\begin{equation}
\boxed{j=j_o}
\end{equation}
Which is nothing but the missing second requirement appearing in the definition of the EPRL spinfoam model. In summary, up to expected ordering ambiguities the homogenous functions
\begin{equation}
|j,m\rangle:=f^{(\beta j,j)}_{j,m}
\end{equation}
of one spinor variable perfectly solve the linear simplicity constraints \eref{simpl}.

These functions do in fact solve the constraint $F_2=0$ strongly, be seen as follows. Consider the quantisation of $F_2$ as introduced in \eref{F2}.
A short moment of reflection reveals this operator maps homogenous functions of degree $(a,b)$ towards those of degree $(a-1,b+1)$. In terms of the $\rho$, $j_o$ description $\rho$ remains invariant but $j_o$ is shifted to $j_o+1$. Notice now that the constraint commutes with the generators of the $SU(2)$ subgroup that leaves $n^{\alpha\bar\alpha}$ invariant:
\begin{equation}
[\widehat{L}_i,\widehat{F}_2]=0.
\end{equation}
By Schur's lemma we thus get
\begin{equation}
\widehat{F}_2f^{(\rho,j_o)}_{j,m}=c(\rho,j_o,j)f^{(\rho,j_o+1)}_{j,m},
\end{equation}
with some $c(\rho,j_o,j)\in\mathbb{C}$. But $j_o=j$ is the lowest spin appearing, therefore we must have that $c(\rho,j_o,j_o)=0$. Hence
\begin{equation}
\widehat{F}_2|j,m\rangle=0.
\end{equation}
A straightforward calculation proves that
\begin{equation}
\big[\widehat{F}_2,\widehat{F}^\dagger_2\big]|j,m\rangle=2j|j,m\rangle.
\end{equation}
Thus
\begin{equation}
\forall j\neq 0:\widehat{F}^\dagger_2|j,m\rangle\neq 0.
\end{equation}
Therefore just one of the $\mathbb{C}$-valued constraint equations  $F_2=0=\bar{F}_2$ is solved strongly, $\widehat{F}_2$ annihilates physical states, but $\widehat{F}^\dagger_2$ as a kind of creation operator maps them to the orthogonal complement of the solution space. This is an important observation, which should be compared with the Gupta--Bleuler formalism. This is done in the conclusion of this paper.

Notice also that the homogenous functions are not normalisable with respect to the $L^2$ norm on $\mathbb{C}^2$. In order to introduce an inner product a complex surface integral may be used. In fact the equation
\begin{equation}
\langle f,f^\prime\rangle=\frac{\I}{2}\int_{\mathrm{P}\mathbb{C}^2}\omega_\alpha\di\omega^\alpha\wedge\bar{\omega}_{\bar\alpha}\di\bar{\omega}^{\bar\alpha}\overline{f(\omega)}f^\prime(\omega)
\end{equation}
introduces the canonical \cite{unrepsl} inner product between homogenous functions of degree $(-j_o-1+\I\rho,j_o-1+\I\rho)$, with respect to which the basis elements $f^{(\rho,j_o)}_{j,m}$ are all orthogonal, and the integration domain denotes the complex projective space $\mathrm{P}\mathbb{C}^2$. 
\section{Discussion and conclusion}
This article consists of two parts. First of all we gave a spinorial decomposition of the truncated phase space on a graph. We attached a twistor to each of the ends of a link. For this construction to work light-like surfaces, that are faces the flux $\Sigma_{IJ}[f]$ of which is null (i.e. $\Sigma_{IJ}\Sigma^{IJ}=0$), had to be removed. Otherwise our spinorial framework were ill-defined. Since loop quantum gravity mostly considers spacelike boundaries this restriction might not be that serious.

Next, we gave an application. We showed that the linear simplicity constraint $\Sigma_{IJ}n^J=0$ decomposes into the two independent conditions $F_1=0$ and $\boldsymbol{\mathsf{M}}=\bar{F}_2F_2=0$, which commute under the Poisson bracket, whereas $F_2$ and $\bar F_2$ don't. In quantum theory only the former can be imposed strongly. We then took the space of unitary irreducible representations of the Lorentz group, and searched for states being annihilated by the corresponding operators. We found the following restrictions on the quantum numbers of the canonical $f^{(\rho,j_o)}_{j,m}$ basis elements:
\begin{equation*}
\rho=\beta j_o,\;\mathrm{and}\;j=j_o.
\end{equation*}
That is the continuous parameter $\rho$ is related to the discrete $j_o$ label, and all representations are in the lowest spin $j=j_o$ appearing. These are the two equations most crucial for the definition of the EPRL vertex amplitude \cite{flppdspinfoam, lorentzvertam, LQGvertexfinite}, derived in a framework completely different from the one the spinfoam model was originally built in. We consider this result to be an important observation strongly supporting the EPRL model.

Finally, using a simple argument we proved that the spinorial functions $|j,m\rangle$ 
lying in the common kernel of $\boldsymbol{\widehat{\mathsf{M}}}$ and $\widehat{F}_1$, are automatically annihilated by $\widehat{F}_2$, but in general $\widehat{F}^\dagger_2 |j,m\rangle\neq 0$. Therefore the pair $\widehat{F}_2, \widehat{F}_2^\dagger$ is imposed weakly, which reminds us of the way Gupta and Bleuler \cite{gupta, bleuler} removed longitudinal photons. But there are of course subtle differences we briefly discuss in the following. 

In QED Gupta \cite{gupta} and Bleuler \cite{bleuler} managed to impose gauge conditions on the electromagnetic 4-potential (e.g. $\Omega=\partial_a A^a=0$). Splitting the gauge fixing function into parts of positive $\Omega^+$ and negative frequency $\Omega^-$, they required physical states to be annihilated by $\Omega^+$. In our case, of course, there is no notion of positive/negative frequency. But the constraints are complex and one may use the complex structure to give another decomposition. Indeed we could quantise the phase space by means of analytic wavefunctions, and impose only the analytic part of the constraint equations thereon. In fact the founders of the twistorial program of loop quantum gravity already performed first steps towards this task for both the Lorentzian and the Euclidean case \cite{twistcons, twistconslor}. Here we did something different. Our wavefunctions $f(\omega)$ are non-analytic functions on $\mathbb{C}^2$, that is the Cauchy--Riemann differential equations 
\begin{equation}
\quad\frac{\partial}{\partial \bar\omega^{\bar\alpha}}f= 0\label{cauchriem}
\end{equation}
do \emph{not} hold. 

An analytic quantisation  \cite{penroserindler2} would then require that the wavefunction depend on both $\omega$ and $\bar\pi$, abd fulfil \eref{cauchriem} also for $\pi$. This would lead us to a different quantisation of the canonical Poisson commutation relations (see again \cite{penroserindler2} on that). One would then find a new solution space for the simplicity constraints, and should ask whether the two results are isomorphic to one another.

Let us also mention a couple of related publications apeared just recently \cite{spezialetwist1, spezialetwist2}. These papers are much more detailed on the classical level, the authors explicitly keep the underlying graph unspecified (here discussions are mostly restricted to one single link), search for a semiclassical meaning of the variables introduced, and---most importantly---perform the reduction induced by the simplicity constraints from $SL(2,\mathbb{C})$ variables down to $SU(2)$. Concerning the classical part our results perfectly match, differences appear on the side of quantum theory. Here we've chosen a \emph{non-analytic} quantisation of the complex phase space (similar to the Landau quantisation of an electron in a 2-dimensional plane $\mathbb{C}\ni z=x+\I y$ perpendicular to a constant magnetic field), whereas the authors of \cite{spezialetwist1, spezialetwist2} seem to prefer an analytic quantisation (similar to the Bargmann quantisation of the 1-dimensional harmonic oszillator in terms of $z=x+\I p$). These differences do definitely deserve further examination. 
\section*{Acknowledgments}
I wish to thank Simone Speziale for proposing the project this article is built on. I am grateful for numerous discussions with Eugenio Bianchi, Simone Speziale, Carlo Rovelli and Johannes Tambornino which I had the opportunity to participate in. In fact they considerably improved this article. 
\appendix
\section{Elementary definitions and basic notation}\label{appA}
\paragraph*{Index conventions.}
Indices $I,J,K,\dots\in\{0,\dots,3\}$ from the middle of the roman alphabet refer to four dimensional Minkowski space $(\mathbb{R}^4,\eta_{IJ})$, their lowercase counterparts run from one to three. The metric signature is $(-,+,+,+)$, and $\epsilon_{0123}=1$ determines the Levi-Civita tensor. A left handed spinor, that is an element $v^\alpha$ of $\mathbb{C}^2$ is decorated with upperstage indices $\alpha,\beta,\dots\in\{0,1\}$, elements $u^{\bar\alpha}$ of the complex conjugate vector space $\bar{\mathbb{C}}^2$ of opposite chirality are commonly marked with either dotted or primed indices, here we use \qq{bar-ed} indices $\bar\alpha,\bar\beta,\dots$ instead. Brackets $(\dots)$ (and equally for $[\dots]$) around a number of indices denote total (anti-)symmetrisation of all intervening symbols.
\newline
\paragraph*{Phase space conventions.}
Consider the symplectic potential $\Theta=p\di q$.
For any two differentiable functions $f,f^\prime$ on phase space, their respective Hamiltonian vector fields define their common Poisson bracket
\begin{equation}
\iota_{\mathfrak{X}_f}\Omega=\di f,\quad\big\{f,f^\prime\big\}=\Omega(\mathfrak{X}_f,\mathfrak{X}_{f^\prime}),
\end{equation}
constructed from the symplectic 2-form $\Omega=\di\Theta$. From this the only non vanishing Poisson bracket between the phase space variables reads
\begin{equation}
\{p,q\}=1.
\end{equation}
\paragraph*{Spinors and the Lorentz group.}
The $SL(2,\mathbb{C})$ group is the universal cover of the group $L_+^\uparrow$ of proper orthochronous Lorentz transformations, in fact the map $\Lambda$ relating one to the other is determined by the defining equation:
\begin{equation}
\eqalign{
\Lambda:SL(2,\mathbb{C})\ni g\mapsto\Lambda(g)\in L_+^\uparrow:\\
\ou{g}{\mu}{\alpha}\ou{\bar{g}}{\bar{\mu}}{\bar{\alpha}}\ou{\sigma}{\alpha\bar{\alpha}}{I}=\ou{\Lambda(g)}{J}{I}\ou{\sigma}{\mu\bar{\mu}}{J}.
}\label{lambdadef}
\end{equation}
The \qq{intertwining} $\ou{\sigma}{\alpha\bar\beta}{I}$ symbols are a basis in the four dimensional vector space of Hermitian $2\times 2$ matrices $X^{\alpha\bar\alpha}$, establish an isomorphism between those and vectors in Minkowski space, and obey for any $X^I\in\mathbb{R}^4$ that:
\begin{equation}
\mathrm{det}\big(X^I\sigma_I\big)=-\eta_{IJ}X^IX^J.
\end{equation}
Note that this implicitly shows $\Lambda(g)$ to be an element of $L_+^\uparrow$. Following the general conventions of \cite{sexlurbantke2} these matrices are chosen to be:
\begin{equation}
\sigma_0=\pmatrix{1 & 0\cr 0 & 1},\quad\sigma_i=i\mbox{-th Pauli matrix}\label{Paulm}.
\end{equation}
The antisymmetric $\epsilon$-tensor, together with its inverse, invariant under the action of unimodular matrices:
\begin{equation}
\eqalign{\forall g\in SL(2,\mathbb{C}):\epsilon_{\mu\nu}\ou{g}{\mu}{\alpha}\ou{g}{\nu}{\beta}=\epsilon_{\alpha\beta},\\
\epsilon^{\mu\nu}:\epsilon^{\mu\alpha}\epsilon_{\nu\alpha}=\delta^\mu_\nu,}
\end{equation}
is used in order to establish an isomorphism between $\mathbb{C}^2$ and its dual vector space:
\begin{eqnarray}
\mathbb{C}^2\ni\omega^\mu&\longmapsto \omega_\mu=\epsilon_{\nu\mu}\omega^\nu\in\big(\mathbb{C}^2\big)^\ast,\\
\big(\mathbb{C}^2\big)^\ast\ni\omega_\mu&\longmapsto \omega^\mu=\epsilon^{\mu\nu}\omega_\nu\in\mathbb{C}^2.
\end{eqnarray}
Here one has to be careful with index positions, particularly illustrated by the identity:
\begin{equation}
\xi_\mu v^\mu=-\xi^\mu v_\mu.
\end{equation}
Our conventions are finally fixed by giving the explicit matrix elements of $\epsilon^{\mu\nu}$ and $\epsilon_{\mu\nu}$:
\begin{equation}
\epsilon_{01}=\epsilon^{01}=1
\end{equation}
The traceless matrices
\begin{equation}
\ou{\big[S_{(IJ)}\big]}{\mu}{\nu}=-\frac{1}{2}\ou{\sigma}{\mu\bar{\alpha}}{K}{\bar{\sigma}}_{\bar{\alpha}\nu L}\delta^K_{[I}\delta^L_{J]},
\end{equation}
are a basis in $\mathfrak{sl}(2,\mathbb{C})$ and establish the isomorphism induced by \eref{lambdadef} between $\mathfrak{so}(1,3)$ and 
$\mathfrak{sl}(2,\mathbb{C})$ via:
\begin{equation}
\Lambda_\ast:\mathfrak{sl}(2,\mathbb{C})\ni\frac{1}{2}S_{(IJ)}\omega^{IJ}\longmapsto\ou{\omega}{I}{J}\in\mathfrak{so}(1,3).\label{lambdaiso}
\end{equation}
These generators correspond to the selfdual sectors of the Lorentz algebra, e.g.
\begin{equation}
S_{(IJ)}\ou{P}{IJ}{MN}=S_{(MN)}.
\end{equation}
Where we have introduced the selfdual projector
\begin{equation}
\ou{P}{IJ}{MN}=\frac{1}{2}\big(\delta^{[I}_M\delta^{J]}_N-\frac{\I}{2}\ou{\epsilon}{IJ}{MN}\big).
\end{equation}
Furthermore for any $\omega\in\mathfrak{so}(1,3)$ we find that:
\begin{equation}
\frac{1}{2}\omega^{IJ}S_{(IJ)}=\tau_i\Big(\frac{1}{2}\uo{\epsilon}{m}{in}\ou{\omega}{m}{n}+\I\ou{\omega}{i}{o}\Big),
\end{equation}
where $\sigma_i=2\I\tau_i$ are the Pauli spin matrices, and for any $\omega\in\mathfrak{so}(1,3)$
\begin{equation}
\omega^i=\frac{1}{2}\uo{\epsilon}{m}{in}\ou{\omega}{m}{n}+\I\ou{\omega}{i}{o}\label{selfcomp}
\end{equation}
denote its selfdual components. Choosing these complex coordinates on $\mathfrak{sl}(2,\mathbb{C})$ calculations generally simplify, in fact the commutation relations between the selfdual generators are nothing but
\begin{equation}
\big[\tau_i,\tau_j\big]=\uo{\epsilon}{ij}{m}\tau_m.
\end{equation}
\paragraph*{Canonical basis.}
The canonical basis vectors $f^{(\rho,j_o)}_{j,m}$ in the $(-j_o-1+\I\rho,j_o+1+\I\rho)$ irreducible unitary representation space of $SL(2,\mathbb{C})$ \cite{ruhl, gelminshap, unrepsl} diagonalise a complete set of commuting observables:
\begin{eqnarray}
\big(\widehat{L}^2-\widehat{K}^2\big)f^{(\rho,j_o)}_{j,m} =\big(j_o^2-1-\rho^2\big)f^{(\rho,j_o)}_{j,m},\\
\widehat{L}_i\widehat{K}^if^{(\rho,j_o)}_{j,m}=-j_o\rho f^{(\rho,j_o)}_{j,m},\\
\widehat{L}^2f^{(\rho,j_o)}_{j,m}=j(j+1)f^{(\rho,j_o)}_{j,m},\\
L_3f^{(\rho,j_o)}_{j,m}=mf^{(\rho,j_o)}_{j,m}.
\end{eqnarray}\label{eigenv}
The operators $\widehat{K}_i$ and $\widehat{L}_i$ are the infinitesimal generators of boosts and rotations.
The label $\rho\in\mathbb{R}$ is continuous but $2 j_o\in\mathbb{N}_0$ is discrete. Each irreducible unitary representation of $SL(2,\mathbb{C})$ automatically induces a representation of the $SU(2)$ subgroup of rotations. The quantum numbers $j=j_o,j_o+1,\dots$ and $m=-j,\dots, j$ refer to the respective $SU(2)$ irreducible subspaces, and $j_o$ is the lowest spin appearing.
\section{Proofs and calculations}\label{appB}
\paragraph*{Twisted rotations.}
Here we prove that the \qq{twisted} spinor transformations defined as in (\ref{twistrota}\emph{-b}) actually leave the group element unchanged. Let us first observe that for any point in our spinorial phase space the requirement $H\neq 0$ allows us to write the $\epsilon$-tensor in terms of phase space variables:
\begin{equation}
\epsilon^{\mu\nu}=\frac{\pi^\mu\omega^\nu-\pi^\nu\omega^\mu}{H}.\label{edecomp}
\end{equation}
The twisted rotations read:
\begin{eqnarray}
\utilde{\omega}^\alpha & \longrightarrow -\frac{\utilde{\pi}^\alpha\pi_\mu+\utilde{\omega}^\alpha\omega_\mu}{H}\ou{G}{\mu}{\nu}\pi^\nu,\\
\utilde{\pi}^\alpha & \longrightarrow +\frac{\utilde{\pi}^\alpha\pi_\mu+\utilde{\omega}^\alpha\omega_\mu}{H}\ou{G}{\mu}{\nu}\omega^\nu.
\end{eqnarray}
Inserting these equations together with (\ref{twistrota}\emph{-b})  into the definition \eref{gdef} of the group element proves the property of invariance:
\begin{eqnarray}
\nonumber\ou{g}{\alpha}{\beta} \longrightarrow&  -\frac{\utilde{\pi}^\alpha\pi_\mu+\utilde{\omega}^\alpha\omega_\mu}{\det G\; H\sqrt{H\utilde{H}}}\ou{G}{\mu}{\nu}\uo{G}{\beta}{\rho}\big(\omega^\nu\pi_\rho-\pi^\nu\omega_\rho\big)\\
&=-\frac{\utilde{\pi}^\alpha\pi_\mu+\utilde{\omega}^\alpha\omega_\mu}{\det G\sqrt{H\utilde{H}}}\ou{G}{\mu}{\nu}\uo{G}{\beta}{\nu}=\ou{g}{\alpha}{\beta},\label{twistrotprf}
\end{eqnarray}
When going from the first towards the second line one needs \eref{edecomp} together with the fact that for any $G\in GL(2,\mathbb{C})$
\begin{equation}
\epsilon_{\mu\nu}\ou{G}{\mu}{\alpha}\ou{G}{\nu}{\beta}=\det G\;\epsilon_{\alpha\beta}.
\end{equation}
\paragraph*{Area matching condition.}
Here we show why the equation $\utilde{\Pi}^{\alpha\beta}=-\ou{g}{\alpha}{\mu}\ou{g}{\beta}{\nu}\Pi^{\mu\nu}$ requires the \qq{area matching} constraint \eref{armatch2} to be fulfilled. Using \eref{gdef} we find:
\begin{eqnarray}
\nonumber\utilde{\Pi}^{\alpha\beta}&\stackrel{!}{=}-\frac{1}{4}\ou{g}{\alpha}{\mu}\ou{g}{\beta}{\nu}(\pi^\mu\omega^\nu+\pi^\nu\omega^\mu)=\\
\nonumber&=-\frac{1}{4H\utilde{H}}(\utilde{\pi}^\alpha\utilde{\omega}^\beta\pi_\mu\omega^\mu\omega_\nu\pi^\nu
 +\utilde{\pi}^\beta\utilde{\omega}^\alpha\omega_\mu\pi^\mu\pi_\nu\omega^\nu)=\\
&=\frac{1}{4}\frac{H}{\utilde{H}}(\utilde{\pi}^\alpha\utilde{\omega}^\beta+\utilde{\pi}^\beta\utilde{\omega}^\alpha).
\end{eqnarray}
Since singular configurations $\utilde{H}\neq 0$ have already been removed, we get the desired constraint
\begin{equation}
C=H-\utilde{H}=0.
\end{equation}
\paragraph*{Poisson bracket of two group elements.}
We are now going to compute the Poisson bracket between two group elements provided the constraint $C=H-\utilde{H}=0$ holds. What we get is this:
\begin{eqnarray}
\fl
\nonumber\big\{\ou{g}{\alpha}{\beta},\ou{g}{\alpha^\prime}{\beta^\prime}\big\}_{\mathbb{P}}\big|_{C=0}   = \bigg\{\frac{\utilde{\pi}^\alpha\pi_\beta+\utilde{\omega}^\alpha\omega_\beta}{\sqrt{H\utilde{H}}},\frac{\utilde{\pi}^{\alpha^\prime}\pi_{\beta^\prime}+\utilde{\omega}^{\alpha^\prime}\omega_{\beta^\prime}}{\sqrt{H\utilde{H}}}\bigg\}_{\mathbb{P}}\bigg|_{C=0}=\\
\fl\nonumber \quad = \frac{1}{H^2}\big[\epsilon^{\alpha\alpha^\prime}(\pi_\beta\omega_{\beta^\prime}+\pi_{\beta^\prime}\omega_\beta)
+\epsilon_{\beta\beta^\prime}(\utilde{\pi}^\alpha\utilde{\omega}^{\alpha^\prime}+\utilde{\pi}^{\alpha^\prime}\utilde{\omega}^\alpha)\big]_{C=0}+\\
\fl\nonumber\qquad-\frac{1}{H^2}\big(\utilde{\pi}^\alpha\pi_\beta-\utilde{\omega}^\alpha\omega_\beta\big)\ou{g}{\alpha^\prime}{\beta^\prime}\big|_{C=0}+
\frac{1}{H^2}\big(\utilde{\pi}^{\alpha^\prime}\pi_{\beta^\prime}-\utilde{\omega}^{\alpha^\prime}\omega_{\beta^\prime}\big)\ou{g}{\alpha}{\beta}\big|_{C=0}=\\
\fl\nonumber\quad=\frac{4}{H^2}\big(\epsilon^{\alpha\alpha^\prime}\Pi_{\beta\beta^\prime}+\epsilon_{\beta\beta^\prime}\utilde{\Pi}^{\alpha\alpha^\prime}\big)_{C=0}
-\frac{2}{H^3}\big(\utilde{\pi}^\alpha\utilde{\omega}^{\alpha^\prime}\pi_\beta\omega_{\beta^\prime}-\utilde{\pi}^{\alpha^\prime}\utilde{\omega}^{\alpha}\pi_{\beta^\prime}\omega_{\beta}\big)_{C=0}=\\
\fl\nonumber\quad=\frac{4}{H^2}\big(\epsilon^{\alpha\alpha^\prime}\Pi_{\beta\beta^\prime}+\epsilon_{\beta\beta^\prime}\utilde{\Pi}^{\alpha\alpha^\prime}\big)_{C=0}
-\frac{1}{H^3}\Big[(\utilde{\pi}^\alpha\utilde{\omega}^{\alpha^\prime}-\utilde{\pi}^{\alpha^\prime}\utilde{\omega}^{\alpha})(\pi_\beta\omega_{\beta^\prime}+\pi_{\beta^\prime}\omega_{\beta})+\\
\fl\qquad+(\utilde{\pi}^\alpha\utilde{\omega}^{\alpha^\prime}+\utilde{\pi}^{\alpha^\prime}\utilde{\omega}^{\alpha})(\pi_\beta\omega_{\beta^\prime}-\pi_{\beta^\prime}\omega_{\beta})\Big]_{C=0}=0.
\end{eqnarray}
Where $\Pi^{\alpha\beta}=4(\pi^\alpha\omega^\beta+\pi^\beta\omega^\alpha)$ and the last equality follows from the decomposition \eref{edecomp} of the $\epsilon$-invariant.
\paragraph*{Master constraint.}
Be $n^I$ a timelike unit vector $n_In^I=-1$, consider the following decomposition of the master constraint:
\begin{eqnarray}
\nonumber\boldsymbol{\mathsf{M}} & = \bar{F}_2F_2=n^{\beta\bar\alpha}n^{\alpha\bar\beta}\omega_\beta\pi_\alpha\bar\pi_{\bar\alpha}\bar\omega_{\bar\beta}=\\
\nonumber&=n^{\beta\bar\alpha}n^{\alpha\bar\beta}\big(\omega_{[\beta}\pi_{\alpha]}+\omega_{(\beta}\pi_{\alpha)}\big)\bar{\pi}_{\bar\alpha}\bar{\omega}_{\bar\beta}=\\
\nonumber&=\frac{1}{2}n^{\beta\bar\alpha}n^{\alpha\bar\beta}\epsilon_{\beta\alpha}\epsilon^{\mu\nu}\omega_\mu\pi_\nu\bar\pi_{\bar\alpha}\bar\omega_{\bar\beta}+
n^{\beta\bar\alpha}n^{\alpha\bar\beta}\omega_{(\beta}\pi_{\alpha)}\bar\pi_{(\bar\alpha}\bar\omega_{\bar\beta)}=\\
&=-\frac{1}{2}\omega^\mu\pi_\mu\bar{\pi}_{\bar\mu}\bar\omega^{\bar\mu}+4n^{\beta\bar\alpha}n^{\alpha\bar\beta}\Pi_{\beta\alpha}\bar\Pi_{\bar\alpha\bar\beta}.\label{Mcons}
\end{eqnarray}
Let us now introduce generators of boosts and rotations:
\begin{equation}
\Pi_{\alpha\beta}=\tau_{\alpha\beta i}\Pi^i=-\frac{1}{2}\tau_{\alpha\beta i}\big(L^i+\I K^i\big).
\end{equation}
We are in the timelike case and can thus always choose a frame such that $n^{\alpha\bar\alpha}={1\;0 \choose 0\;1}$. With respect to this gauge the Pauli matrices become Hermitian, this in turn implies:
\begin{equation}
\ou{\tau}{\alpha}{\beta i}=-n^{\alpha\bar\beta}n_{\beta\bar\alpha}\ou{\bar\tau}{\bar\alpha}{\bar\beta i}.
\end{equation}
Using $-2\mathrm{Tr}(\tau_i\tau_j)=\delta_{ij}$ we finally get:
\begin{equation}
\boldsymbol{\mathsf{M}}=-\frac{1}{2}\bar\omega^{\bar\mu}\bar{\pi}_{\bar\mu}\pi_\mu\omega^\mu-\frac{1}{2}\big(L^2-K^2\big)+L^2.
\end{equation}
Where $L^2=L_iL_j\delta^{ij}$ and $K^2=K_iK_j\delta^{ij}$.
\section*{References}

\end{document}